\documentclass{cs19proc}
\usepackage{amsmath,amssymb,bm,titlesec}

\newcommand{\vv}{\mathbf{v}}

\newcommand{\vB}{\mathbf{B}}

\newcommand{\vJ}{\mathbf{J}}

\newcommand{\bomega}{\boldsymbol{\Omega}}

\newcommand{\grad}{\boldsymbol{\nabla}}
\newcommand{\bcdot}{\boldsymbol{\cdot}}
\newcommand{\dvg}{\boldsymbol{\nabla}\!\bcdot\!}
\newcommand{\curl}{\boldsymbol{\nabla}\!\!\boldsymbol{\times}\!\!}
\newcommand{\cnabla}{\!\boldsymbol{\cdot}\!\boldsymbol{\nabla}}
\newcommand{\cross}{\!\!\boldsymbol{\times}\!\!}

\editors{G.~A. Feiden}
\publisher{Zenodo}
\conference{The 19th Cambridge Workshop on Cool Stars, Stellar Systems, and the Sun}
\conferencedate{2016}

\title{Simple Scaling Relationships for Stellar Dynamos}
\author{Kyle Augustson$^1$, St\'{e}phane Mathis$^{1}$, Allan Sacha Brun$^{1}$}

\affiliation{
  $^1$Laboratoire AIM Paris-Saclay, CEA/DRF -- CNRS -- Universit\'{e} Paris Diderot, IRFU/SAp Centre de Saclay, F-91191 Gif-sur-Yvette Cedex, France}

\shorttitle{Dynamo Scaling}
\shortauthors{Augustson}

\abs{This paper provides a brief overview of dynamo scaling relationships for the degree of
  equipartition between magnetic and kinetic energies. Three basic approaches are adopted to explore
  these scaling relationships, with a first look at two simple models: one assuming magnetostrophy
  and another that includes the effects of inertia. Next, a third scaling relationship is derived
  that utilizes the assumptions that the dynamo possesses two integral spatial scales and that it is
  driven by the balance of buoyancy work and ohmic dissipation as studied in \citet{davidson13}. The
  results of which are then compared to a suite of convective dynamo simulations that possess a
  fully convective domain with a weak density stratification and that captured the behavior of the
  resulting dynamo for a range of convective Rossby numbers \citep{augustson16}.}

\begin{document}

\maketitle

\section{Introduction}

A precise rubric for predicting the nature of the saturated state of turbulent convective dynamos
remains as elusive as tracing individual convective eddies: they can be identified for a time, but
they eventually are lost in the tumult. Nevertheless, one can hope to approximate the shifting
nature of those dynamos. The effects of astrophysical dynamos can be detected at the surface and in
the environment of a given magnetically-active object, such as stars
\citep[e.g.,][]{christensen09,donati09,donati11,brun15}. One direct way to approximate the dynamics
occurring within such an object is to conduct laboratory experiment with fluids that have some
equivalent global properties, while observing their response to controllable parameters, such as the
strength of thermal forcing or rotation rate. In those cases, all the observable scales of the
system can be accounted for, from the global or driving scale to the dissipation scales. In
practice, this has proven to be quite difficult when trying to mimic geophysical or astrophysical
dynamos, but they are still fruitful endeavors \citep[e.g.,][]{gailitis99,laguerre08,spence09}.
However, recent experiments with liquid gallium have shown that magnetostrophic states, where the
Coriolis force balances the Lorentz force, seem to be optimal for heat transport \citep{king15},
which is interesting given the strong likelihood that many astrophysical dynamos are in such a
state. Another method is to simulate a portion of those experiments, but these numerical simulations
are limited in the scales they can capture: either an attempt is made to resolve a portion of the
scales in the inertial range to down to the physical dissipation scale
\citep[e.g.,][]{mininni09a,mininni09b,brandenburg14}, or an attempt is made to approximate the
equations of motion for the global scales while modeling the effects of the unresolved dynamical
scales \citep[e.g.,][]{gilman83,brun04,christensen06,strugarek16}.

These varying approaches to gathering data about the inner workings of convective dynamos provide a touchstone for
thought experiments. Further, one can attempt to identify a few regimes for which some global-scale aspects of those
dynamos might be estimated with only a knowledge of the basic parameters of the system. The following questions are
examples of such parametric dependencies: how the magnetic energy contained in the system may change with a
modified level of turbulence (or stronger driving), or how does the ratio of the dissipative length scales impact that
energy, or how does rotation influence it?  Establishing the global-parameter scalings of convective dynamos,
particularly with stellar mass and rotation rate, is useful given that they provide an order of magnitude approximation
of the magnetic field strengths generated within the convection zones of stars as they evolve from the pre-main-sequence
to a terminal phase. This in turn permits the placement of better constraints upon transport processes, such as those
for elements and angular momentum, most of which occur over structurally-significant evolutionary timescales.

\section{Fundamental Equations}

In the hunt for a simple set of algebraic equations to describe the basic processes at work, it is
useful to consider the following set of MHD equations:

\vspace{-0.25truein}
\begin{center}
  \begin{align}
    &\frac{\partial\rho}{\partial t} = -\dvg{\left(\rho\vv\right)}, \nonumber \\
    &\frac{\partial\vv}{\partial t} = -\left(\right.\!\!\vv\cnabla{\!\left.\right)\!\vv}
    - 2\bomega\cross\vv - \frac{\grad{P}}{\rho} - \grad{\Phi_{\mathrm{eff}}}
    +\frac{\vJ\cross\vB}{c\rho} +\frac{\grad{\!\bcdot\sigma}}{\rho}, \label{eqn:eom} \\
    &\frac{\partial\vB}{\partial t} = \curl{\left[\vv\cross\vB-\eta\vJ/c\right]}, \nonumber \\
    &\grad{\bcdot\vB}=0, \nonumber \\
    &\frac{\partial E}{\partial t} = -\dvg{\left[\left(E + P - \sigma\right)\vv +\mathbf{q}\right]} + \rho\epsilon, \nonumber \\
    &\frac{\partial \rho s}{\partial t} = -\dvg{\left[\rho s \vv\right]} +\frac{1}{T}\left[\frac{4\pi\eta}{c^2}\vJ\cdot\vJ +
                                         \sigma:\nabla\vv + \rho\epsilon - \dvg{\mathbf{q}}\right], \nonumber
  \end{align}
\end{center}

\noindent where $\vv$ is the velocity, $\vB$ is the magnetic field, $\rho$ is the density, $P$ the
pressure, and $s$ is the entropy per unit mass. Moreover, the following variables are also defined
as $\Phi_{\mathrm{eff}} = \Phi + \lambda^2\Omega^2/2$, $\Phi$ is the gravitational potential,
$\Omega$ is the rotation rate of the frame, $\lambda$ is the distance from the axis of rotation, the
current is $\vJ = c\curl{\vB}/4\pi$, with $c$ being the speed of light, $\sigma$ is the viscous
stress tensor, $\mathbf{q} = -\kappa \nabla T$, $\kappa$ is the thermal diffusivity and $\epsilon$
is an internal heating rate per unit mass that is due to some prescribed exoergic process (e.g.,
chemical, nuclear, or otherwise). The total energy is $E =\rho v^2/2 + B^2/8\pi +
\rho\Phi_{\mathrm{eff}} + \rho e$, where $e$ is the internal energy per unit mass.

Since only the most basic scaling behavior of the stellar system is sought, the following
assumptions are made: the total energy of the system is conserved and second the system is in a
time-steady, nonlinearly-saturated equilibrium. The first assumption leads to the elimination of the
volume-integrated energy equation.

\section{Scaling of Magnetic and Kinetic Energies} \label{sec:forcescaling}

One way to assess what the scaling behavior of the magnetic and kinetic energies in a dynamo is to
find the balance of forces acting in the system when in a quasi-statistically steady, but nonlinear
regime. To get a feeling for the basic balances, consider Equation (\ref{eqn:eom}) first for a
slowly rotating system, where the buoyancy, Coriolis, pressure, and viscous forces are neglected.
Such a simple force balance involves the Lorentz and inertial force as

\vspace{-0.25truein}
\begin{center}
\begin{align}
  \rho\vv\cdot\nabla\vv \approx \frac{1}{4\pi}\left(\curl{\vB}\right)\cross\vB , \label{eqn:forcebalance}
\end{align}
\end{center}

\noindent where $\vv$ is the velocity, $\vB$ is the magnetic field, and $\rho$ is the density.
Suppose that the flow and magnetic field vary with some characteristic length scale $\ell$.
Therefore, this balance yields a magnetic field strength in equipartition with the kinetic energy
contained in the convection, such as

\vspace{-0.25truein}
\begin{center}
\begin{align}
  B^2_{\mathrm{eq}} \approx 4\pi\rho \mathrm{v_{rms}^2}.
\end{align}
\end{center}

Convective flows often possess distributions of length scales and speeds that are peaked near a
single characteristic value. One simple method to estimate these quantities is to divine that the
energy containing flows have roughly the same length scale as the depth of the convection zone and
that the speed of the flows is directly related to the rate of energy injection (given here by the
stellar luminosity) and inversely proportional to the density of the medium into which that energy
is being injected \citep{augustson12}. The latter is encapsulated as $v_{rms} \propto
\left(2L/\rho_{\mathrm{CZ}}\right)^{1/3}$, where $\rho_{\mathrm{CZ}}$ is the average density in the
convection zone. However, such a mixing-length velocity prescription only provides an order of
magnitude estimate as the precise level of equipartition depends sensitively upon the dynamics
\citep[e.g.,][]{yadav16}. Since stars are often rotating fairly rapidly, taking for instance young
low-mass stars and most intermediate and high-mass stars, their dynamos may reach a
quasi-magnetostrophic state wherein the Coriolis acceleration also plays a significant part in
balancing the Lorentz force.  Such a balance has been addressed and discussed at length in
\citet{christensen10} and \citet{brun15} for instance.  To again have a zeroth-order suggestion of
the behavior of more rapidly-rotating convective dynamos, note that there are three other forces at
work in addition to the Coriolis, inertial, and Lorentz forces, namely the forces resulting from
pressure gradients, buoyancy, and viscous diffusion. Assuming that some fraction ($1-\beta$, for
$0<\beta<1$) of the inertial force accounts for these remaining forces, it can be seen that

\vspace{-0.25truein}
\begin{center}
\begin{align}
  &\beta\rho\vv\cdot\nabla\vv + 2\rho\bomega_0\cross\vv \approx \frac{1}{4\pi}\left(\curl{\vB}\right)\cross\vB, \nonumber \\
  &\implies \frac{\beta}{\ell}\rho \mathrm{v_{rms}^2} + 2\rho \mathrm{v_{rms}} \Omega_0 \approx \frac{B^2}{4\pi \ell}, \nonumber \\
  &\implies \frac{B^2}{8\pi} \approx \frac{1}{2}\rho \mathrm{v_{rms}^2}\left(\beta + 2 \ell \Omega_0/\mathrm{v}\right), \nonumber\\
  &\implies \frac{\mathrm{ME}}{\mathrm{KE}} \approx \beta + \mathrm{Ro}^{-1}, \label{eqn:scaling}
\end{align}
\end{center}

\noindent with $\bomega_0 = \Omega_0 \hat{e}_z$ the rotation rate of the reference frame and with
$\mathrm{Ro}$ the convective Rossby number (hereafter denoted as only the Rossby number).

As demonstrated in \citet{augustson16}, Equation (\ref{eqn:scaling}) may hold for a subset of convective dynamos,
wherein the ratio of the total magnetic energy (ME) to the kinetic energy (KE) depends on the inverse Rossby number and
a constant offset.  The constant is sensitive to details of the dynamics and, in some circumstances, it may also be
influenced by the Rossby number. In any case, convective dynamos are sensitive to the degree of the rotational
constraint on the convection, as it has a direct impact on the intrinsic ability of the convection to generate a
sustained dynamo.  Yet, even in the absence of rotation, there appears to be dynamo action that gives rise to a minimum
magnetic energy state in the case of sufficient levels of turbulence.  Hence, there is a bridge between two dynamo
regimes: the equipartition slowly rotating dynamos and the rapidly rotating magnetostrophic regime, where
$\mathrm{ME/KE}\propto \mathrm{Ro}^{-1}$.  For low Rossby numbers, or large rotation rates, it is even possible that the
dynamo can reach superequipartition states where $\mathrm{ME/KE}>1$. Indeed, it may be much greater than unity, as is
expected for the Earth's dynamo (see Figure 6 of \citet{roberts13}).\\

To better characterize the force balance without directly resorting to the parameterization above,
consider again Equation (\ref{eqn:eom}) but this time taking its curl, wherein one can see that

\vspace{-0.25truein}
\begin{center}
  \begin{align}
    \frac{\partial\boldsymbol{\omega}}{\partial t} &= \curl{\left[\vv\cross\boldsymbol{\omega_P}+\frac{1}{\rho}\left(\frac{\vJ\cross\vB}{c}+\dvg{\sigma}-\grad P\right)\right]},
  \end{align}
\end{center}

\noindent where $\boldsymbol{\omega} = \curl{\vv}$ and $\boldsymbol{\omega_P}=2\bomega+\boldsymbol{\omega}$.  

Taking the dot product of this equation with $\boldsymbol{\omega}$ gives rise to the equation for
the evolution of the enstrophy.  Integrating that equation over the volume of the convective domain
and over a reasonable number of dynamical times such that the system is statistically steady yields

\vspace{-0.25truein} 
\begin{center}
  \begin{align}
    &\!\!\int\!\! d\mathbf{S}\!\bcdot\!\left[\vv\cross\boldsymbol{\omega_P}+\frac{1}{\rho}\left(\frac{\vJ\cross\vB}{c}+\dvg{\sigma}-\grad P\right)\right]\cross\boldsymbol{\omega} \nonumber \\
    &\!\!+\!\!\int\!\! dV \left(\curl{\boldsymbol{\omega}}\right)\!\bcdot\!\left[\vv\cross\boldsymbol{\omega_P}+\frac{1}{\rho}\left(\frac{\vJ\cross\vB}{c}+\dvg{\sigma}-\grad P\right)\right]\!=\!0.
  \end{align}
\end{center}

\noindent If no enstrophy is lost through the boundaries of the convective domains, then the surface
integral vanishes, leaving

\vspace{-0.25truein} 
\begin{center}
  \begin{align}
    \!\!\int\!\! dV \left(\curl{\boldsymbol{\omega}}\right)\!\bcdot\!\left[\vv\cross\boldsymbol{\omega_P}+\frac{1}{\rho}\left(\frac{\vJ\cross\vB}{c}+\dvg{\sigma}-\grad P\right)\right]=0.
  \end{align}
\end{center}

\noindent This assumption effectively means that magnetic stellar winds will not be part of this
scaling analysis.  For timescales consistent with the dynamical timescales of the dynamos considered
here, this is a reasonably valid assumption. Then, since $\boldsymbol{\nabla\!\times\!\omega}$ is
not everywhere zero, the terms in square brackets must be zero, which implies that

\vspace{-0.25truein} 
\begin{center}
  \begin{align}
    \rho\vv\cross\boldsymbol{\omega_P}+\frac{\vJ\cross\vB}{c}+\dvg{\sigma}-\grad P=0.
  \end{align}
\end{center}

\noindent Taking the curl of this equation eliminates the pressure contribution and gives

\vspace{-0.25truein}
\begin{center}
  \begin{align}
    \curl{\left[\rho\vv\cross\boldsymbol{\omega}+2\rho\vv\cross\bomega+\frac{\vJ\cross\vB}{c}+\dvg{\sigma}\right]}=0.\label{eqn:curlforce}
  \end{align}
\end{center}

\noindent This is the primary force balance, being between inertial, Coriolis, Lorentz, and viscous forces.  Taking
fiducial values for the parameters, and scaling the derivatives as the inverse of a characteristic length scale $\ell$,
the scaling relationship for the above equation yields

\vspace{-0.25truein} 
\begin{center}
  \begin{align}
    \rho \mathrm{v_{rms}^2}/\ell^2 + 2\rho \mathrm{v_{rms}}\Omega_0/\ell+ B^2/4\pi\ell^2 + \rho\nu \mathrm{v_{rms}}/\ell^3 \approx 0,
  \end{align}
\end{center}

\noindent which when divided through by $\rho\mathrm{v_{rms}^2}/\ell^2$ gives

\vspace{-0.25truein} 
\begin{center}
  \begin{align}
    \mathrm{ME/KE} \propto 1 + \mathrm{Re}^{-1} + \mathrm{Ro}^{-1}.
  \end{align}
\end{center}

\noindent The Reynolds number used in the above equation is taken to be $\mathrm{Re} = \mathrm{v_{rms}}\ell/\nu$.  Since
the curl is taken, the approximation for the pressure gradient and buoyancy terms employed in Equation
(\ref{eqn:scaling}) is eliminated from the force balance.  However, the leading term of this scaling relationship is
found to be less than unity, at least when assessed through simulations.  Hence, it should be replaced with a parameter
to account for dynamos that are subequipartition, leaving the following

\vspace{-0.25truein} 
\begin{center}
  \begin{align}
    \mathrm{ME/KE} \propto \beta(\mathrm{Ro,Re}) + \mathrm{Ro}^{-1}, \label{eqn:forcescaling}
  \end{align}
\end{center}

\noindent where $\beta(\mathrm{Ro,Re})$ is unknown a priori as it depends upon the intrinsic ability of the non-rotating
system to generate magnetic fields, which in turn depends upon the specific details of the system such as the boundary
conditions and geometry of the convection zone.

\section{An Alternative Approach to Scaling Relationships for ME/KE}\label{sec:alternative}

An alternative approach to building a scaling relationship for the ratio of magnetic to kinetic energy considers the
balances established in generating entropy, kinetic energy, and magnetic energy. To begin, note that the evolution of
the magnetic energy is

\vspace{-0.25truein}
\begin{center}
  \begin{align}
    \frac{\partial}{\partial t}\left(\frac{\vB^2}{8\pi}\right) &=
    \frac{\vB}{4\pi}\!\bcdot\!\curl{\left[\vv\cross\vB-\frac{\eta}{c}\vJ\right]}, \nonumber \\
    &=-\frac{1}{c}\left[\dvg{\left(\eta\vJ\cross\vB\right)}+\frac{4\pi\eta}{c}\vJ^2+\vv\!\bcdot\!\left(\vJ\cross\vB\right)\right].
  \end{align}
\end{center}

\noindent If this equation is averaged over many dynamical times $\tau$, when it is in a quasi-steady state, and if it
is integrated over the volume of the convective region, it yields

\vspace{-0.25truein}
\begin{center}
  \begin{align}
    &\!\!\int\!\!\frac{dt}{\tau} dV\left[\frac{4\pi\eta}{c}\vJ^2+\vv\!\bcdot\!\left(\vJ\cross\vB\right)\right] = -\!\!\int\!\!\frac{dt}{\tau} d\mathbf{S}\!\bcdot\!\left(\eta\vJ\cross\vB\right). \label{eqn:intme}
  \end{align}
\end{center}

\noindent The Lorentz force ($\vJ\cross\vB$) can vanish at the boundaries of the convective region
for an appropriate choice of boundary conditions.  As an example, if the magnetic field satisfies a
potential field boundary condition, then it is zero. Or if the field is force-free (e.g.,
$\vJ\propto\vB$), then it is also zero.  Supposing that this is the case, then one has that the
average Lorentz work ($\int\!\!dt/\tau dV\vv\!\bcdot\!\left(\vJ\cross\vB\right)$) is equal to the
average Joule heating $\epsilon_{\eta}=4\pi\!\!\int\!\! dt/\tau dV\eta\vJ^2/c$.  This is an
important point that is often brushed aside in astrophysics, as it shows that the nature of the
convection and magnetic field structures are directly impacted by the form of the resistive
dissipation.  Hence, the use of numerical dissipation schemes could yield unexpected results.

In a fashion similar to that used for the magnetic energy evolution above, one can find that the
kinetic energy evolves as

\vspace{-0.25truein}
\begin{center}
  \begin{align}
    \frac{1}{2}\frac{\partial\rho\vv^2}{\partial t} &=
    -\dvg{\left[\left(\rho\vv^2/2+P-\sigma\right)\vv\right]} + P\dvg{\vv} - \sigma\!:\!\grad\vv \nonumber \\
    &-\rho\vv\cnabla\Phi_{\mathrm{eff}} + \frac{\vv}{c}\!\bcdot\!\left(\vJ\cross\vB\right).
  \end{align}
\end{center}

\noindent To eliminate the pressure, the total internal energy must also be added to the system as

\vspace{-0.25truein}
\begin{center}
  \begin{align}
    &\frac{\partial}{\partial t}\left[\rho\vv^2/2 + \vB^2/8\pi + \rho e\right] = \nonumber \\
    &-\dvg{\left[\left(\rho\vv^2/2 + \rho e + P - \sigma\right)\vv - \frac{\eta}{c}\vJ\cross\vB\right]} \nonumber \\
    &-\rho\vv\cnabla{\Phi_{\mathrm{eff}}} -\frac{4\pi\eta}{c^2}\vJ^2 -\sigma\!:\!\grad\vv + \rho\epsilon-\dvg{\mathbf{q}}.
  \end{align}
\end{center}

\noindent One can also consider the time-averaged and volume-integrated evolution equation for this
energy equation, which yields

\vspace{-0.25truein}
\begin{center}
  \begin{align}
    &\!\!\int\!\! \frac{dt}{\tau} dV\left[\rho\epsilon-\dvg{\mathbf{q}}
      -\rho\vv\cnabla\Phi_{\mathrm{eff}}-\frac{4\pi\eta}{c^2}\vJ^2-\sigma:\grad\vv\right]=\nonumber \\
    &\!\!\int\!\! \frac{dt}{\tau}
    d\mathbf{S}\!\bcdot\!\left[\left(\rho\vv^2/2 + \rho e + P - \sigma\right)\vv - \frac{\eta}{c}\vJ\cross\vB\right].
  \end{align}
\end{center}

\noindent The surface integral is zero if there are no outflows or net torque from the Lorentz force
at the domain boundaries, implying the following:

\vspace{-0.25truein}
\begin{center}
  \begin{align}
    &\!\!\int\!\! \frac{dt}{\tau} dV\left[\rho\epsilon-\dvg{\mathbf{q}}
      -\rho\vv\cnabla\Phi_{\mathrm{eff}}-\frac{4\pi\eta}{c^2}\vJ^2-\sigma:\grad\vv\right]=0. \label{eqn:nrg}
  \end{align}
\end{center}

\noindent Note that $\int\! dV\rho\epsilon = L(r)$, where $L$ is the total luminosity of the star at
a given radius $r$ for a spherically symmetric heating.  Likewise, the radiative luminosity of the
star is given by $\int\! dV \grad\!\!\bcdot\!{\mathbf{q}} = L_r(r) = -4\pi r^2\kappa\partial
T/\partial r$ for the spherically-symmetric component of the temperature field, which should be
dominant. For the case of stars that are on the main-sequence, there are three configurations of
their primary regions of convection: either a convective core for high mass stars, a convective
envelope for lower mass stars, or both for F-type stars.  In all these cases, one can assume that
the region of integration is over the entire convection zone and so the volume-integrated luminosity
will be the nuclear luminosity, or the current total luminosity $L_{*}$.  Furthermore, the
inwardly-directed radiative luminosity will be nearly, but not exactly, equal in magnitude to
$L_{*}$.  The reason being that the thermal evolution of the system is a largely passive response to
the changes in the nuclear burning rates.  Because the nuclear luminosity is slowly increasing along
the main-sequence, $L_r$ will always lag behind $L_{*}$ due to the time required for thermal
diffusion to modify the thermal gradient. From Equation (\ref{eqn:nrg}), one can see that

\vspace{-0.25truein}
\begin{center}
  \begin{align}
   L_r + \epsilon_\nu + \epsilon_\eta - L_{*} = -\int \frac{dt}{\tau} dV \rho\vv\cnabla\Phi_{\mathrm{eff}},
  \end{align}
\end{center}

\noindent where $\epsilon_\nu$ and $\epsilon_\eta$ are the positive-definite, time-averaged,
volume-integrated dissipation rates due to viscosity and resistivity, respectively. Thus, the rate
of buoyancy work is directly proportional to the mismatch of the two luminosities and the rates of
viscous and resistive dissipation, implying that the latter result from the former. Assuming that
$L_* \approx L_r$ and following \citet{brandenburg14}, the ratio of the dissipation rates can be
described as

\vspace{-0.25truein}
\begin{center}
  \begin{align}
    \epsilon_\nu/\epsilon_\eta = k \mathrm{Pm}^n,
  \end{align}
\end{center}

\noindent where $\mathrm{Pm}=\nu/\eta$ is the magnetic Prandtl number, and where $k$ and $n$ could be determined from a
suite of direct numerical simulations. In particular, when kinetic helicity is injected at the driving scale,
\citet{brandenburg14} found that $k=7/10$ and $n=2/3$. Note that their results have also considered rotating driven
turbulence and found that this scaling is effectively independent of the rotation rate, and thus the Rossby number.
However, other studies indicate that there may be a stronger rotational influence \citep{plunian10}.  Since there is
ambiguity in that scaling, let $k(\mathrm{Ro})$ be an unknown function of the Rossby number. Subsequently, the buoyancy
work $W_B$ per unit mass can be described as

\vspace{-0.25truein}
\begin{center}
  \begin{align}
    W_B &= -\frac{\int \frac{dt}{\tau} dV \rho\vv\cnabla\Phi_{\mathrm{eff}}}{\int \frac{dt}{\tau}dV\rho} = \left(1+k(\mathrm{Ro}) \mathrm{Pm}^{2/3}\right)\epsilon_\eta/M.
  \end{align}
\end{center}

\noindent where $M$ is the mass in the integrated volume.

Now, returning to the time-averaged curl of the momentum equation, though neglecting the viscous and
inertial terms, one can find

\vspace{-0.25truein}
\begin{center}
  \begin{align}
    \curl{\left[2\rho\vv\cross\bomega+\frac{1}{c}\vJ\cross\vB\right]}+\grad\rho\cross\mathbf{g}_{\mathrm{eff}}=0, \label{eqn:curlmom}
  \end{align}
\end{center}

\noindent where $\mathbf{g}_{\mathrm{eff}}=-\grad\Phi_{\mathrm{eff}}$.

This provides the basis of a scaling relationship. Following \citet{davidson13}, the Rossby number
is assumed to be small enough so that the flow becomes roughly columnar and moderately aligned with
the rotation axis. In such a case, there are two integral length scales: one parallel to the
rotation axis $\ell_{\|}$ and another perpendicular to it $\ell_{\bot}$, with
$\ell_{\bot}<\ell_{\|}$. Here, unlike \citet{davidson13}, the density stratification is
retained. So, density perturbations rather than temperature perturbations are contained in the
buoyancy work integral and the force balance below. The full density can be retained in the integral
and in the scaling given that the gradient of the mean density is parallel to
$\mathbf{g}_{\mathrm{eff}}$. So, their cross product vanishes, leaving only the product of the
velocity and gradients of the density perturbations. Assuming further that the magnetic energy
density per unit mass scales only with $\ell_{\|}$ and $W_B$, unit consistency requires that
$B^2/(4\pi\rho) \propto F\left(\ell_{\|}, W_B\right) \approx \ell_{\|}^{2/3} W_B^{2/3}$, where $W_B$
is the rate of buoyancy work per unit mass as above. Therefore, given Equation (\ref{eqn:curlmom})
and assuming that each term is of the same order of magnitude, the basic proportionality is

\vspace{-0.25truein}
\begin{center}
  \begin{align}
    &\rho\bomega\cnabla\vv\approx\grad\rho\cross\mathbf{g}_{\mathrm{eff}}\approx\curl{\left(\vJ\cross\vB\right)}/c \nonumber \\
    \implies &\frac{\Omega_0 \mathrm{v_{rms}}}{\ell_{\|}} \approx \frac{g}{\ell_{\bot}} \approx \frac{B^2}{4\pi\rho\ell_{\bot}^2}. \label{eqn:forcebal}
  \end{align}
\end{center}

Thus, comparing the curl of the Lorentz force to the curl of the Coriolis force, one has

\vspace{-0.25truein}
\begin{center}
  \begin{align}
    \Omega_0 \mathrm{v_{rms}}/\ell_{\|} \approx B^2/4\pi\rho\ell_{\bot}^2 \approx \ell_{\|}^{2/3}W_B^{2/3}/\ell_{\bot}^2, 
  \end{align}
\end{center}

\noindent and moreover it can be shown that an estimate of the rms velocity is

\vspace{-0.25truein}
\begin{center}
  \begin{align}
    \mathrm{v_{rms}}\approx \ell_{\|}^{5/3}W_B^{2/3}/\ell_{\bot}^2\Omega_0.
  \end{align}
\end{center}

\noindent Within the context of such estimates, the buoyancy term can be written as

\vspace{-0.25truein}
\begin{center}
  \begin{align}
    W_B = \frac{\int \frac{dt}{\tau} dV \rho\vv\bcdot\mathbf{g}_{\mathrm{eff}}}{\int \frac{dt}{\tau}dV\rho} \approx g\mathrm{v_{rms}},
  \end{align}
\end{center}

\noindent which implies that the estimated magnitude for the curl of the buoyancy force in Equation
(\ref{eqn:forcebal}) is $g/\ell_{\bot} \approx W_B/(\mathrm{v_{rms}}\ell_{\bot})$. Then, it is easily
seen that

\vspace{-0.25truein}
\begin{center}
  \begin{align}
    \ell_{\bot}/\ell_{\|} \approx W_B/(\Omega_0\mathrm{v_{rms}^2}).
  \end{align}
\end{center}

\noindent So, the ratio of integral length scales should vary as

\vspace{-0.25truein}
\begin{center}
  \begin{align}
    \frac{\ell_{\bot}}{\ell_{\|}} \approx \frac{W_B^{1/9}}{\Omega_0^{1/3} \ell_{\|}^{2/9}} = \left(\frac{W_B}{\Omega_0^3\ell_{\|}^2}\right)^{1/9}.
  \end{align}
\end{center}

\noindent Likewise the ratio of magnetic to kinetic energy then scales as

\vspace{-0.25truein}
\begin{center}
  \begin{align}
    \mathrm{\frac{ME}{KE}} \approx \frac{B^2}{8\pi\left(1/2\rho\mathrm{v_{rms}^2}\right)} \approx \left(\frac{W_B}{\Omega_0^3\ell_{\|}^2}\right)^{-2/9}.
  \end{align}
\end{center}

\noindent To make a comparison to the earlier results in \S\ref{sec:forcescaling}, consider that the
Rossby number is defined as

\vspace{-0.25truein}
\begin{center}
  \begin{align}
    \mathrm{Ro} = \frac{\mathrm{v_{rms}}}{\Omega_0\ell_{\|}}\approx\left(\frac{W_B}{\Omega_0^3\ell_{\|}^2}\right)^{4/9}, \label{eqn:rossby}
  \end{align}
\end{center}

\noindent which implies that

\vspace{-0.25truein}
\begin{center}
  \begin{align}
    \mathrm{\frac{ME}{KE}} \propto \mathrm{Ro}^{-1/2}. \label{eqn:bouyancyscaling}
  \end{align}
\end{center}

\noindent However, if one takes into account the scaling of the buoyancy work with Rossby number and
magnetic Prandtl number, Equation (\ref{eqn:rossby}) becomes an implicit relationship for the Rossby
number that is indeterminate for large magnetic Prandtl number.

\begin{figure}[t]
  \centering
  \includegraphics[width=0.95\linewidth]{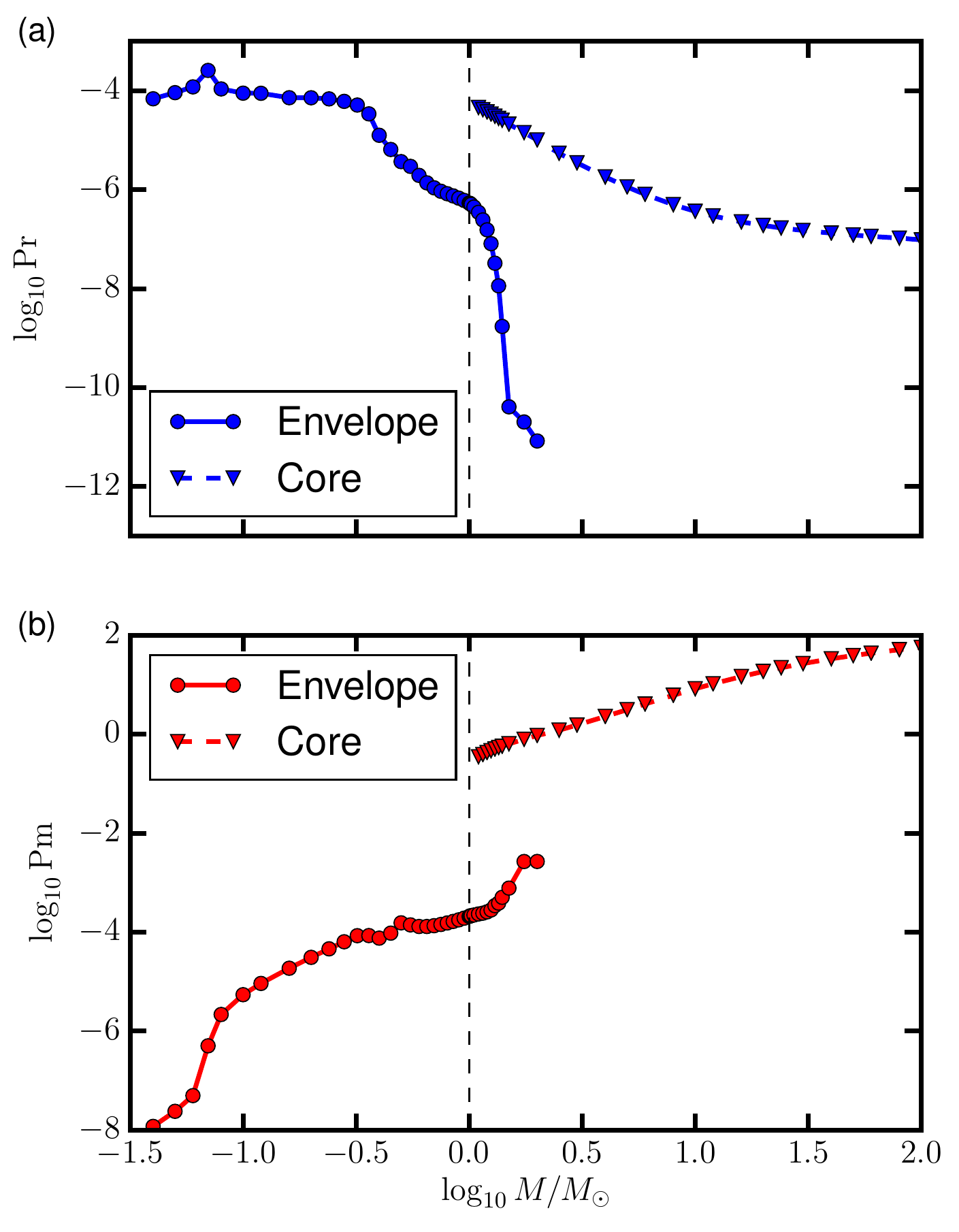}
  \caption{The average thermal Prandtl number $\mathrm{Pr}$ (a) and average magnetic Prandtl number $\mathrm{Pm}$ (b)
    for stars with masses between 0.03 and 100~$M_{\odot}$, with the average being taken over substantial convection
    zones.  The zone being averaged is indicated by triangles for a convective core and circles for a convective
    envelope. }\vspace{-0.25truein}
  \label{fig:prandtl}
\end{figure}

\section{Comparison of Scaling Relationships}

Since the scaling described in the previous section has a magnetic Prandtl number dependence, it is useful to quantify
the specific regimes in which each of the three different models are most applicable.  To do so, consider a
fully-resolved convective dynamo and its associated dynamics, wherein the dissipation of the energy injected into the
system is governed by the molecular values of the diffusivities.  Using Braginskii plasma diffusivities
\citep{braginskii65} and a set of MESA stellar models with a solar-like metallicity to obtain the density and
temperature profiles \citep{paxton11}, one can define the average expected molecular magnetic Prandtl number in either
the convective core of a massive star or the convective envelope of a lower mass star.  First, note that the molecular
values of the diffusive coefficients in the Navier-Stokes MHD equations are given by

\vspace{-0.25truein}
\begin{center}
  \begin{align}
    \nu = 3.2 10^{-5} \frac{T_e^{5/2}}{\rho \Lambda}, \\
    \kappa_{\mathrm{cond}} = 1.8 10^{-6} \frac{T_e^{5/2}}{\rho\Lambda}, \\
    \eta = 1.2 10^6 \Lambda T_e^{-3/2},
  \end{align}
\end{center}

\noindent where $\nu$ is the kinematic viscosity, $\eta$ is the magnetic diffusivity, $\kappa_{\mathrm{cond}}$ is the
electron thermal diffusivity, $\kappa_{\mathrm{rad}}$ is the radiative thermal diffusivity, $T_e$ is the electron
temperature in electron volts, and $\Lambda$ is the Coulomb logarithm.  These quantities are computed under the
assumptions that charge neutrality holds and that the temperatures are not excessively high, so that the Coulomb
logarithm is well-defined. In the charge-neutral regime, the magnetic diffusivity happens to be density independent
because the electron collision time scales as the inverse power of ion density and the conductivity is proportional to
the electron density times the electron collision time. Therefore, the thermal Prandtl number $\mathrm{Pr}$ and magnetic
Prandtl number $\mathrm{Pm}$ scale as

\vspace{-0.25truein}
\begin{center}
  \begin{align}
    \mathrm{Pr} = \frac{\nu}{(\kappa_{\mathrm{cond}}+\kappa_{\mathrm{rad}})}, \\
    \mathrm{Pm} = 2.7 10^{-11} \frac{T_e^{4}}{\rho \Lambda^2}. 
  \end{align}
\end{center}

\noindent In most cases, the radiative diffusivity is several orders of magnitude larger than the electron conductivity
and it thus dominates the thermal Prandtl number.

The above prescription for the Prandtl numbers has been applied to MESA models of stars near the zero-age main-sequence
in the mass range between 0.03 and 100 $M_{\odot}$.  The resulting convective-zone-averaged Prandtl numbers are shown in
Figure (\ref{fig:prandtl}), where it is clear that all stars possess convective regions with a low thermal Prandtl
number, whereas one can find two regimes of magnetic Prandtl number. The existence of these two regimes is directly
related to where the convective region is located. For massive stars with convective cores, the temperature and density
averaged over the convective volume are quite high when compared to lower-mass stars with a convective envelope. Such a
high temperature leads to a large magnetic Prandtl number.  The dichotomy in magnetic Prandtl number implies that there
may be two fundamentally different kinds of convective dynamo action in low-mass versus high-mass stars.

\begin{figure}[!t]
  \centering
  \includegraphics[width=0.95\linewidth]{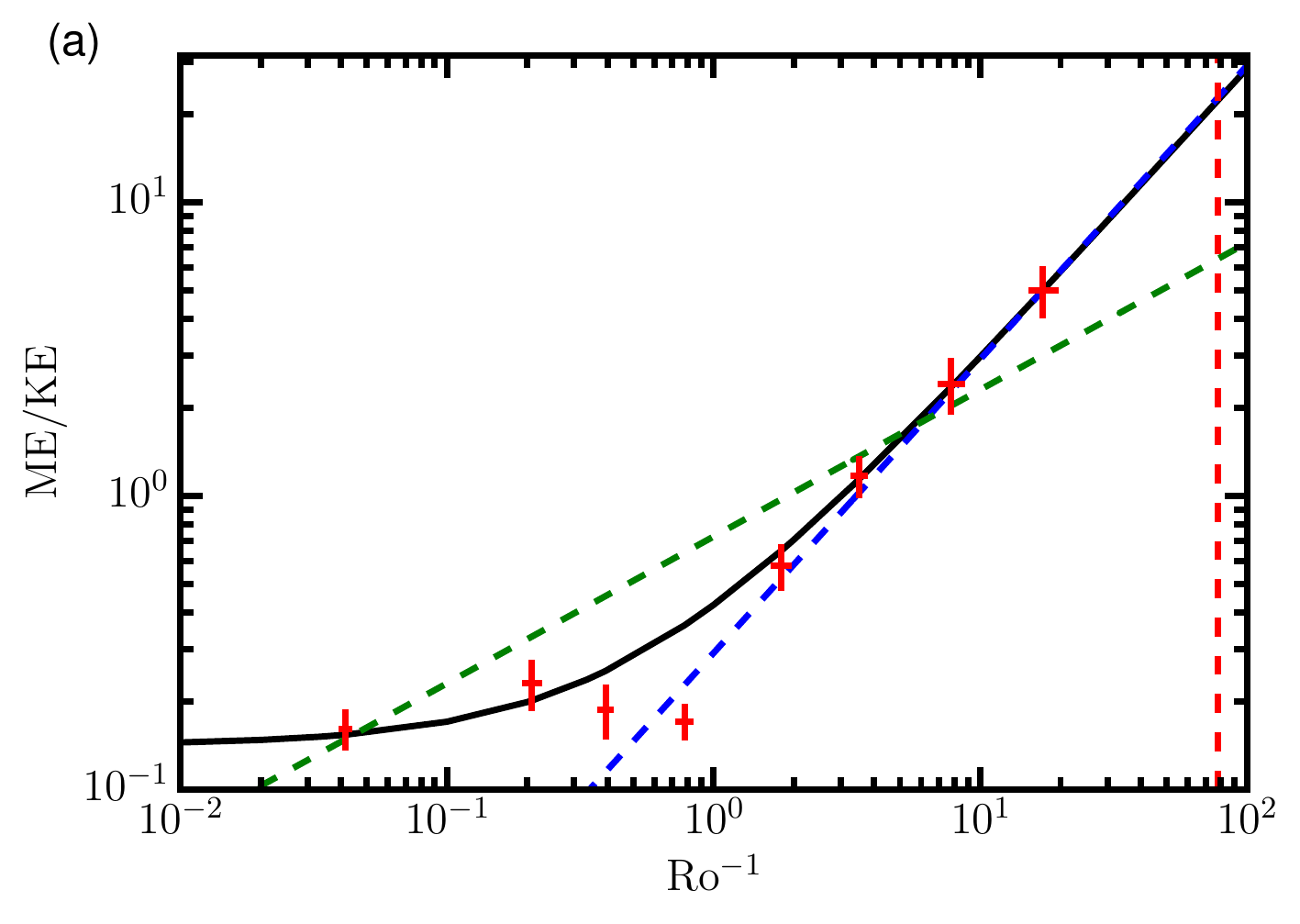}
  \caption{The scaling of the ratio of magnetic to kinetic energy ($\mathrm{ME/KE}$) with inverse
    Rossby number ($\mathrm{Ro}^{-1}$). The black curve indicates the scaling defined in Equation
    (\ref{eqn:forcescaling}), with $\beta=0.5$. The blue dashed line is for magnetostrophy,
    e.g. $\beta=0$.  The green dashed line is that for buoyancy-work-limited dynamo scaling, given
    in Equation (\ref{eqn:bouyancyscaling}). The red dashed line indicates the critical Rossby
    number of the star, corresponding to its rotational breakup velocity. The uncertainty of the
    measured Rossby number and energy ratio that arises from temporal variations are indicated by
    the size of the cross for each data point.}
  \label{fig:scaling}\vspace{-0.25truein}
\end{figure}

\begin{figure}[ht]
  \centering
  \includegraphics[width=0.95\linewidth]{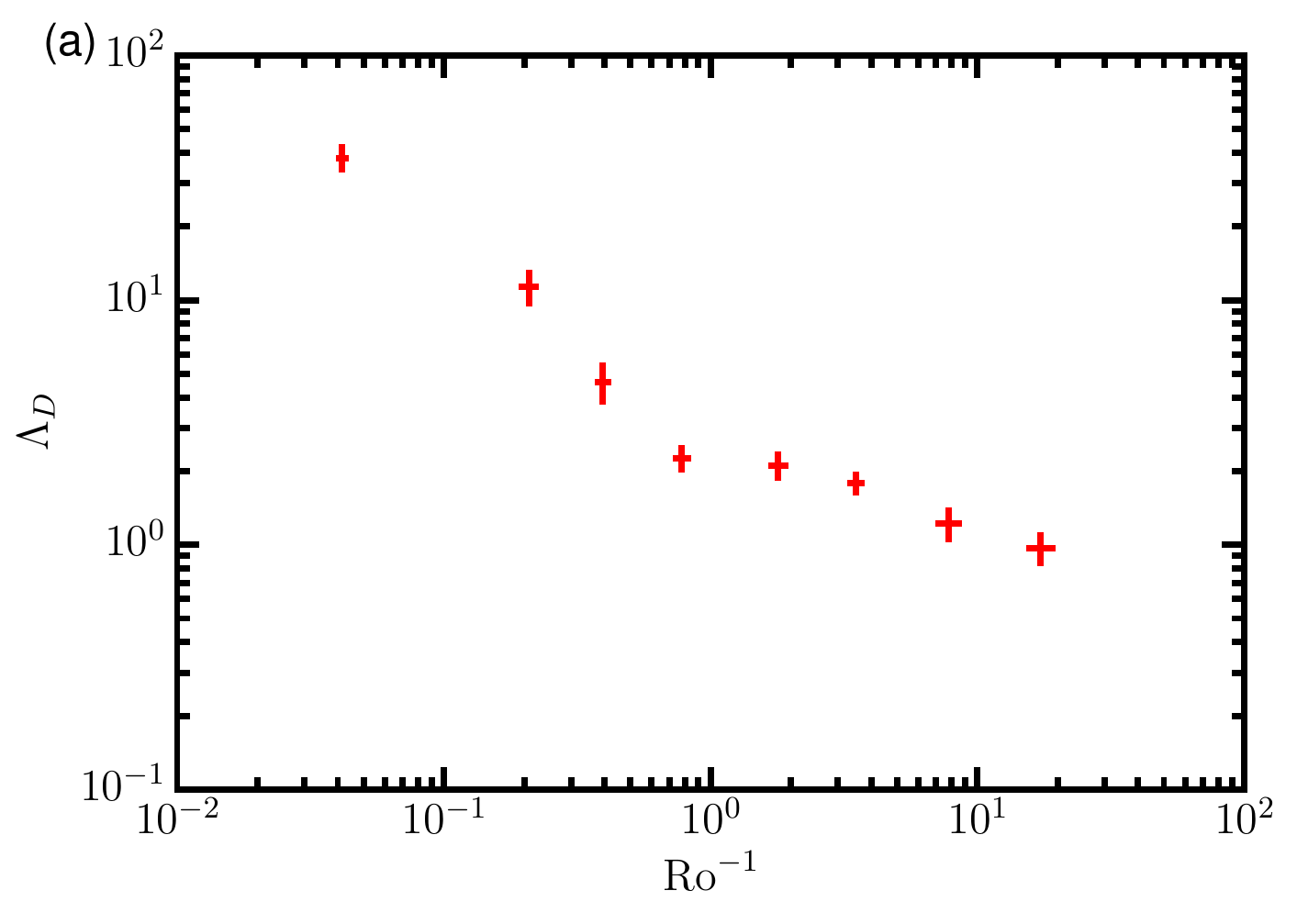}
  \caption{The scaling of the dynamic Elsasser number ($\Lambda_D$) with inverse Rossby number
    ($\mathrm{Ro}^{-1}$). The uncertainty of the measured Rossby number and dynamic Elsasser number that
    arises from temporal variations are indicated by the size of the cross for each data point.}\vspace{-0.25truein}
  \label{fig:elsasser}
\end{figure}

To compare these three schemes, consider the data for the evolution of a set of MHD simulations using the Anelastic
Spherical Harmonic code presented in \citet{augustson16}. These simulations approximate the convective core dynamo that
likely exists within massive stars. In such 10~$M_{\odot}$ stars, the average $\mathrm{Pm}$ is about four throughout the
core, and it drops to roughly $1/10$ close to the stellar surface in the radiative exterior.  Therefore, the convective
core can be considered as a large magnetic Prandtl number dynamo. What makes such $\mathrm{Pm}$ regimes interesting is
that they are a numerically accessible, but still astrophysically relevant, dynamo.  Moreover, these convective cores
represent an astrophysical dynamo in which large-eddy simulations can more easily capture the hierarchy of relevant
diffusive timescales. This is especially so given that the density contrast across the core is generally small, meaning
that one can simplify the problem to being Boussinesq without the loss of too much physical fidelity.  Also, the heat
generation from nuclear burning processes deep within the core and the transition to radiative cooling nearer the
radiative zone are smoothly distributed in radius. So, there are no inherent difficulties with resolving internal
boundary layers.

Within the context of the simulations exhibited in \citet{augustson16}, the rotation rates employed lead to nearly three
decades of coverage in Rossby number, as shown in Figure (\ref{fig:scaling}). In that figure, the force-based scaling
derived in \S\ref{sec:forcescaling} and given by Equation \ref{eqn:forcescaling} is depicted by the black curve, which
does a reasonable job of describing the nature of the superequipartition state for a given Rossby number. Note, however,
that the constant of proportionality has been determined using the data itself.  This is true also of the other two
scaling laws shown in Figure (\ref{fig:scaling}). Surprisingly, the scaling law derived in \S\ref{sec:alternative} does
not capture the behavior of this set of dynamos very well, in contrast to the many dynamo simulations and data shown in
\citet{christensen09} and \citet{christensen10} for which it performs well.  This could be related to the fact that for
large $\mathrm{Pm}$ the buoyancy work potentially has an additional $\mathrm{Ro}$ dependence.

These simulated convective core dynamos appear to enter a magnetostrophic regime for the four cases with the lowest
average Rossby number.  For emphasis, the magnetostrophic scaling regime is denoted by the dashed blue line in Figure
(\ref{fig:scaling}).  This transition to the magnetostrophy is further evidenced in Figure (\ref{fig:elsasser}), which
shows the dynamic Elsasser number

\vspace{-0.25truein}
\begin{center}
  \begin{align}
    \Lambda_D = \mathrm{B_{rms}}^2/(8\pi\rho_0\Omega_0\mathrm{v_{rms}}\ell),
  \end{align}
\end{center}

\noindent where $\ell$ is the typical length scale of the current density $\vJ$. As a point of reference, when
$\Lambda_D$ tends toward unity the balance between the Lorentz and the Coriolis forces also approaches unity, which
indicates that the dynamo is nearly magnetostrophic.

\section{Conclusions}

This conference proceeding hopefully has shed some light on the existence of two kinds of astrophysical dynamos, and
provided scaling relationships for the level of the partitioning of magnetic energy and kinetic energy. Particularly,
there may be a shift in the kind of dynamo action taking place within stars that possess a convective core and those
that possess an exterior convective envelope. The scaling relationship between the magnetic and kinetic energies of such
convective dynamos in turn provide an estimate of the rms magnetic field strength in terms of the local rms velocity and
density at a particular depth in a convective zone.

Two such scaling relationships appear to be the most applicable to the simulations carried out in \citet{augustson16}:
one for the high magnetic Prandtl number regime and another for the low magnetic Prandtl number regime.  Within the
context of the large magnetic Prandtl number systems, the magnetic energy of the system scales as the kinetic energy
multiplied by an expression that depends the inverse Rossby number plus an offset, which in turn depends upon the
details of the non-rotating system (e.g., on Reynolds, Rayleigh, and Prandtl numbers). For low magnetic Prandtl number
and fairly rapidly rotating systems, such as the geodynamo and rapidly rotating stars, another scaling law that relies
upon an energetic balance of buoyancy work and magnetic dissipation (as well as a force balance between the buoyancy,
Coriolis, and Lorentz forces) may be more applicable. When focused on in detail, this yields a magnetic energy that
scales as the kinetic energy multiplied by the inverse square root of the convective Rossby number.  Such a scaling
relationship has been shown to be fairly robust \citep{davidson13}.

Yet, more work is needed to establish more robust scaling relationships that cover a greater range in both magnetic
Prandtl number and Rossby number. Likewise, numerical experiments should explore a larger range of Reynolds number and
level of supercriticality. Indeed, as in \citet{yadav16}, some authors have already attempted to examine such an
increased range of parameters for the geodynamo. Nevertheless, to be more broadly applicable in stellar physics, there
is a need to find scaling relationships that can bridge both the low and high magnetic Prandtl number regimes that are
shown to exist within main-sequence stars. The authors are currently working toward this goal, as will be presented in
an upcoming paper.

\section*{Acknowledgments}

K.~C. Augustson and S. Mathis acknowledge support from the ERC SPIRE 647383 grant. A.~S. Brun
acknowledges funding by ERC STARS2 207430 grant, INSU/PNST, CNES Solar Orbiter, PLATO and GOLF
grants, and the FP7 SpaceInn 312844 grant.

\bibliographystyle{cs19proc}
\bibliography{scaling}

\end{document}